**Near-field Imaging of Optical Resonances in Silicon Metasurfaces Using Photoelectron Microscopy**

Alex Boehm[1], Sylvain D. Gennaro[1,2], Chloe F. Doiron[1,2], Thomas E. Beechem[3], Michael B. Sinclair[1], Igal Brener[1,2], Raktim Sarma[1,2], Taisuke Ohta[1]*

[1]Sandia National Laboratories, Albuquerque, NM 87185, United States

[2]Center for Integrated Nanotechnologies, Albuquerque, NM 87185, United States

[3]School of Mechanical Engineering and Birck Nanotechnology Center, Purdue University, West Lafayette, IN 47907, United States

Corresponding author: tohta@sandia.gov

**ABSTRACT:**

Precise control of light-matter interactions at the nanoscale lies at the heart of nanophotonics. Experimental examination at this length scale is challenging, however, since the corresponding electromagnetic near-field is often confined within volumes below the resolution of conventional optical microscopy. In semiconductor nanophotonics electromagnetic fields are further restricted within the confines of individual subwavelength resonators, limiting access to critical light-matter interactions in these structures. In this work, we demonstrate that photoelectron emission microscopy (PEEM) can be used for polarization resolved near-field spectroscopy and imaging of electromagnetic resonances supported by broken-symmetry silicon metasurfaces. We find that the photoemission results, enabled through an *in-situ* potassium surface layer, are consistent with full-wave simulations and far-field reflectance measurements across visible and near-infrared wavelengths. In addition, we uncover a polarization dependent evolution of collective resonances near the metasurface array edge taking advantage of the far-field excitation and full-field imaging of PEEM. Here, we deduce that coupling between eight resonators or more establishes the collective excitations of this metasurface. All told, we demonstrate that the high-spatial resolution hyperspectral imaging and far-field illumination of PEEM can be leveraged for the metrology of collective, non-local, optical resonances in semiconductor nanophotonic structures.

A recurring desire within nanophotonics is to engineer structures that confine light within a volume smaller than the wavelength of light. Such tight confinement of electromagnetic fields, along with the Purcell effect enhancing the photonic density of states, can greatly intensify light-matter interactions. Resonant structures such as plasmonic antennas,[1,2,3] metasurfaces,[4,5,6] or photonic crystals,[7,8] are known for their ability to manipulate electromagnetic fields efficiently within ultrasmall light-matter interaction volumes, leading to applications in sensing,[9] imaging,[10] holography,[11] nonlinear[12] and quantum optics.[13,14] While miniaturization has enabled control of the light-matter interactions with unprecedented precision, it has made metrology of such photonic structures more challenging, as the spatial resolution of conventional optical microscopy is inadequate for examination of the spatial field distribution.

In recent years, the development of semiconductor nanophotonic structures has granted access to enhanced light-matter interactions strong enough to elicit non-local (*i.e.*, collective) phenomena and non-linear optical responses.[15,16] Light-matter interaction is, in general, polarization dependent due to anisotropic nature of the material response or dielectric permittivity, and in semiconductor metasurfaces these characteristics are frequently enhanced through utilization of asymmetric structures. In this context, it is crucial that nanophotonics systems are investigated using optical illumination with well-defined resolutions in momentum and frequency. Conventional near-field microscopy using electron beam excitations (*e.g.*, cathodoluminescence microscopy and electron energy loss microscopy[17,18,19,20,21]) are limited to indirect examinations of polarization dependent optical phenomena. Whereas for scanning probe-type near-field microscopy,[22,23,24] insertion of a proximal probe in the process of light illumination, or detection, obscures the well-defined polarization of light, in addition to inadvertently impacting the near-field interaction by the proximal probe itself. An approach to overcome such shortcomings is to employ far-field illumination, in which basic characteristics of light (*e.g.*, polarization, energy, and incident geometry) can be controlled within the scope of nanoscale imaging. One such technique that meets these criteria is photoemission electron microscopy (PEEM).

For more than a decade PEEM has been employed to image the near-field distributions of metallic plasmonic nanostructures[25,26,27,28,29,30,31] and, to a smaller extent, photonic resonances in a conductive oxide[32] and excited states in organic molecules.[33,34] However, one must acknowledge that fundamental differences between high-index semiconductor metasurfaces and their plasmonic counterparts make near-field imaging of the former non-trivial. Namely, that semiconductor metasurfaces typically confine electromagnetic fields within the volume of the individual resonators as opposed to plasmonic nanostructures, where electromagnetic fields are evanescent and concentrated at the metal-vacuum or metal-dielectric interface. Nonetheless, PEEM has been utilized in a few semiconductor nanostructure studies to examine localized absorption[35], multiphoton electron excitation, and hot electron dynamics.[36,37] Altogether, the geometric constraints on electromagnetic fields, compounded with additional technical challenges that arise from narrower resonance linewidths (higher *Q*-factor), generally lower energy resonances (longer excitation wavelengths), and higher likelihood of sample charging have posed a significant barrier towards the utilization of PEEM for semiconductor metasurfaces. Each of these challenges are addressed throughout the work presented here.

In this work, we demonstrate polarization resolved spectroscopy and high-resolution imaging of the near-field distribution of optical resonances in a silicon (Si) metasurface using PEEM. The Si metasurface is designed with broken rotational symmetry granting access to quasi-bound states in the continuum (quasi-BICs) that exhibit tight confinement of electromagnetic fields within the body of the resonator, strong polarization dependence, and non-local resonances rendering it an ideal archetype to examine the near-field imaging and spectroscopy of semiconductor nanophotonic structures.[38,39,40] Two-photon photoemission under near-infrared excitation, achieved via a sub-monolayer of potassium (K) deposited *in-situ,* enables nonlinear mapping of the near-field distribution of optical resonances by capturing photoelectron images under different wavelengths and polarizations. We show how the optical resonances evolve at the nanoscale as a function of spatial location within the metasurface providing a detailed account of both the individual optical resonances within a given resonator and the emergence of their collective behavior in aggregate.[15] These findings establish that the versatile hyperspectral near-field imaging capabilities of PEEM can facilitate detailed characterization of both local and non-local (*i.e.*, collective) photonic resonances in semiconductor photonic structures.

**RESULTS & DISCUSSIONS:**

Figure 1a depicts the working concept of photoelectron emission microscopy (PEEM), in which a far-field tunable light source optically excites a sample generating photoelectrons that are collected and redirected onto a 2D imaging detector while preserving their spatial origin. Photoelectron imaging can capture near-field distributions of photonic resonances because photoelectron yield scales with electromagnetic field intensity, meaning that regions where electric fields are more concentrated will produce more photoelectrons, thus generating an image contrast.[41] A photoelectron intensity image therefore reveals the spatial distribution of electromagnetic fields within the sample. This "photon in, electron out" approach offers numerous advantages: far-field excitation with light, akin to typical optical experimental conditions, provides fine control over the polarization, wavelength, incident angle of light, and is able to excite large nanostructures simultaneously[42,43] while electron imaging maintains a high spatial resolution given the small de-Broglie wavelength of the photoelectrons.[44] Multiple photoelectron intensity images can be sequentially acquired while scanning one of the far-field excitation conditions, which is then compiled into a "spectral hypercube" data set. When the varied condition is the excitation wavelength, for example, the photoelectron yield spectra, analogous to far-field optical absorption spectra, can be extracted at each pixel of the PEEM spectral hypercube *post eventum* [45] (see Methods section and Supplementary note in the Supporting Information for more information about the data acquisition and processing).

Photoemission requires that the energy gained by an electron from absorbed photons exceeds the work function of the electron's host material, as illustrated in Figure 1b. However, in semiconductor metasurfaces designed for optical frequencies [from 3 eV (400 nm) to 1.2 eV (1 μm)],[46] the energies of the optical resonances are well below the work function of the resonator material. For this reason, we rely here on multiphoton absorption processes at the optical resonance energies to generate photoelectrons for PEEM imaging, which has been utilized previously with plasmonic nanostructures.[28,42,43,47] In order to increase the efficiency of this nonlinear process we deposit a sub-monolayer of potassium (K) ) onto the metasurface *in-situ* to reduce the work function to ~2.7 eV (see Figure 1b and Supplementary Figure S1 for detail), thereby shifting the threshold for two-photon photoemission into the near-IR [~920 nm (~1.35 eV)]. Photoelectron intensity images generated in this regime can be regarded as nonlinear maps of the near-field distribution of the optical resonances because the intensity of two-photon photoemission is proportional to the square of the electromagnetic field intensity.[25,28,47] The utilization of polarized far-field optical excitation and a large illumination spot (tens of microns, see Supporting Information) with high spatial resolution and full field imaging enables PEEM to uniquely access collective and delocalized photonic resonances supported by a metasurface by imaging many resonators simultaneously.

To showcase this approach, we designed and fabricated a Si metasurface with broken symmetry nanoresonators arranged in periodic arrays (Figure 1c). The metasurface is isolated from a gold-coated Si substrate by a 50 nm dielectric layer stack and coated by a thin (10nm) $TiO_2$ layer to prevent sample charging during PEEM measurements (see Sample fabrication in the Supporting Information, and Supplementary Figure S2). This type of metasurface design exhibits narrowband quasi-BIC optical modes[38,39,40] that are enabled through symmetry breaking operations that permit outcoupling of symmetry-protected BICs to the far-field.[48,49] Spectrally, they appear as sharp resonance peaks in the far-field reflectance measurements. We note that the $TiO_2$ layer and sub-monolayer of K do perturb the optical properties of the metasurfaces to a minor extent, which will be described later. However, the quasi-BIC optical resonances supported by the metasurfaces are preserved. Due to their high-quality factors at normal incidence excitation resulting in significant electromagnetic field concentrations,[50] quasi-BICs are expected to result in significant enhancements of the photoemission process. Moreover, these modes are sensitive to input polarization[51] and their resonances are *collective modes*: *i.e.*, they depend on the location of the resonator within the entire resonator array that are being excited.[15] Thus, these types of quasi-BICs supported by the Si-metasurface are ideally suited for investigation with PEEM.

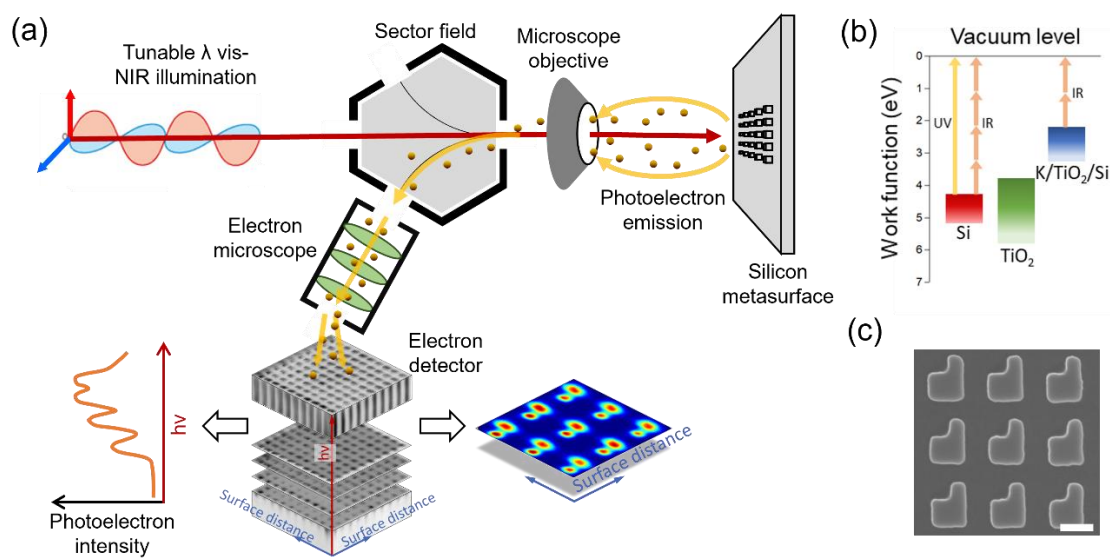

Figure 1: Schematic of the photoelectron emission microscope (PEEM). A Ti:Sapphire oscillator (for two-photon photoelectron excitation) and deep UV laser (for topographical imaging, not depicted) are coupled to a photoelectron microscope. The wavelength tunability of the Ti:Sapphire oscillator allows for acquiring two-dimensional photoelectron yield map as a function of the incident light wavelength. (b) Energy diagram of various multiphoton processes for photoemission and materials' work functions involved in this work. (c) Scanning electron micrograph showing a few individual Si resonators that make up the larger metasurface, the scale bar is 250 nm.

To determine the near-field and far-field optical response of the Si-metasurface, we first performed full-wave numerical simulations using far-field plane wave excitation. The simulated far-field reflectance spectra predict a series of modes spanning the visible to near-IR range for the fabricated Si metasurfaces that exhibit strong polarization dependences. Plotted as 1-reflectance (dashed line) in Figure 2b, these spectra show the different responses for normal incidence excitation at the four polarization orientations (0°, 45°, 90°, and 135°) depicted in Figure 2a. Within these spectra, the two lowest energy modes correspond to quasi-BICs originating from out-of-plane magnetic and electric dipoles, respectively, while additional higher order modes extend towards larger energies. Importantly, each of the four incident polarization orientations is predicted to elicit a distinct spectral profile, where the relative intensity of the modes varies greatly while retaining roughly the same spectral position, providing an ideal system for comparison across far-field and PEEM techniques.

Experimental far-field reflectance spectra in Figure 2c exhibit five resonances spanning the visible to near-IR spectral range, which can be mapped to the modes predicted from simulations. These five resonances peaks are labelled from left to right with letters A-E and are fit with a Lorentzian peak shape using least-squares fitting for easier visualization. We note that the reflectance spectra are acquired from a Si metasurface with a 10 nm $TiO_2$ layer, but without K. To map these resonance peaks to the modes predicted in the simulated spectra, the latter is convoluted with a Gaussian function to represent experimental broadening that originates from imperfect dimensionality of individual resonators, slanted sidewalls, and dimensional variation between the resonator units, as well as instrumental broadening from off-normal incidence optical illumination. The full width at half maximum (FWHM) of the Gaussian function was set to the minimum value (FWHM = 17.5 meV, ~9 nm) required to achieve the same number of final peaks (five) and the resulting broadened simulated spectra are depicted as solid lines in Figure 2b. From here, the five broadened simulated peaks were assigned one-to-one with the experimental resonance peaks starting from the lowest energy peak using the A to E labels.

Following this scheme, we find good matching between simulation and experiment for the energy, spacing, and polarization response of the resonance peaks. In the experimental reflectance spectra, resonances are blue shifted by an average of 50 meV (~25 nm across this wavelength range) however the total energy separation (A to E) remains constant (245 meV versus 275 meV for simulated and far-field reflectance, respectively) supporting these mode designations. Further corroboration of these assignments is found in the polarization response, where the spectral profiles for the experimentally acquired reflectance spectra exhibit clear polarization dependencies that match well with the predictions. For example, the A-resonance peak, which is mapped to the lowest energy peak from simulation, consisting of a single, well isolated mode (out-of-plane magnetic dipole), exhibits the precise polarization dependency predicted from simulation with strong coupling to the far-field at 45° and 90° polarization orientations, but significant suppression at 0° and 135°. Additional polarization responses predicted from simulation can be observed for the other resonance peaks, however they are less apparent as the B- and C- resonances consist of multiple overlapping modes, and the highest energy (D- and E-) resonances, due to their larger quality factors and therefore field enhancements, are more susceptible to losses. One clear discrepancy is the suppression of the B-resonance at 45° which is not predicted from the simulated spectra, as this same discrepancy is noted in the photoelectron yield spectra, less attention is paid in the following discussions.

Photoelectron yield spectra spanning the same visible to near-IR wavelength range captures the four lowest energy resonances (A to D) while retaining the wavelength and polarization responses noted in simulated and far-field reflectance spectra. Photoelectron yield spectra are extracted after *in-situ* K deposition from a series of photoelectron images collected sequentially while varying the near-IR excitation wavelength at a fixed polarization orientation (Figure 2d). Photoelectron yield, or intensity, is integrated over all electron kinetic energies. More information on PEEM data processing is provided in Supplementary note, Analysis of PEEM data. We find that the photoelectron yield spectra correlate well with the simulated and experimental spectra, exhibiting the majority of anticipated wavelength and polarization dependences. For instance, at a 135° polarization orientation, we observe an enhancement of the B- and D- resonances and suppression of the A- and C-resonances, consistent with spectral profiles for the far-field reflectance (Figure 2c) and simulation (Figure 2b). Likewise, at a 90° incident polarization orientation all resonances are strongly expressed, which is again expected. Like the far-field reflectance measurements, discrepancies from the simulated spectra are primarily attributed to limitations in fabrication tolerance and a component of the optical excitation that is off-normal incidence, in this case arising from tilt of the sample for alignment with the electron optics (< 1°). Interestingly, at a 45° polarization orientation, the B resonance is much weaker than the neighboring A and C resonances, as noted previously in the far-field reflectance measurements, but a clear deviation from the simulated spectra. This consistency between experimental methods suggests that photoelectron yield imaging appropriately captures the Si metasurface response under an equivalent far-field optical excitation.

There are some notable differences between the photoelectron yield spectra and reflectance spectra. First, the peak positions in the photoelectron yield spectra are blue shifted by ~35 meV, or ~2% of the excitation energy. This slight blue shift is expected due to the static charge accumulation at the surface of the $TiO_2$ layer from K deposition. An atomic layer of K is known to transfer electrons to a substrate on the order of $10^{13}$ cm$^{-2}$ (corresponding to $10^{19-20}$ cm$^{-3}$) which explains the significant reduction in work function achieved.[52] Such high near-surface electron accumulation would also lower the refractive index, and hence induce sizable blue shift of the resonant mode frequencies.[53] This concept is verified through simulation with and without a K surface layer (see Supplementary Figure S3). This blue shift also explains the absence of the E-resonance in the photoelectron yield spectra as it now lies outside the attainable wavelength range of the light source. We also note some qualitative differences in the peak shape of the photoelectron spectra with respect to the measured reflectance spectra, namely broadening of the respective linewidths (from a 20 – 30 meV FWHM in reflectance to a 20 – 50 meV in PEEM) and an asymmetric peak shape tailing towards lower energies. We postulate that the surface charge accumulation also contributes to these changes in the peak shape, as the electrons accumulated near the surface would induce band bending in the $TiO_2$ film leading to a gradual reduction of the blue shift as a function of distance from the surface. Such asymmetry is captured in the spectral fitting by employing two Lorentzian peaks in the resonances that show obvious tailing. The larger photoelectron linewidths may also, in part, come from the broad spectral width of the Ti:Sapphire laser (8 – 14 meV across the excitation wavelength range used) and the variability in the size and shape of the resonators. Overall, we find a reasonable agreement between measured far-field absorption and the photoelectron yield spectra, affirming the polarization resolved near-field spectroscopy capabilities of PEEM for semiconductor metasurfaces.

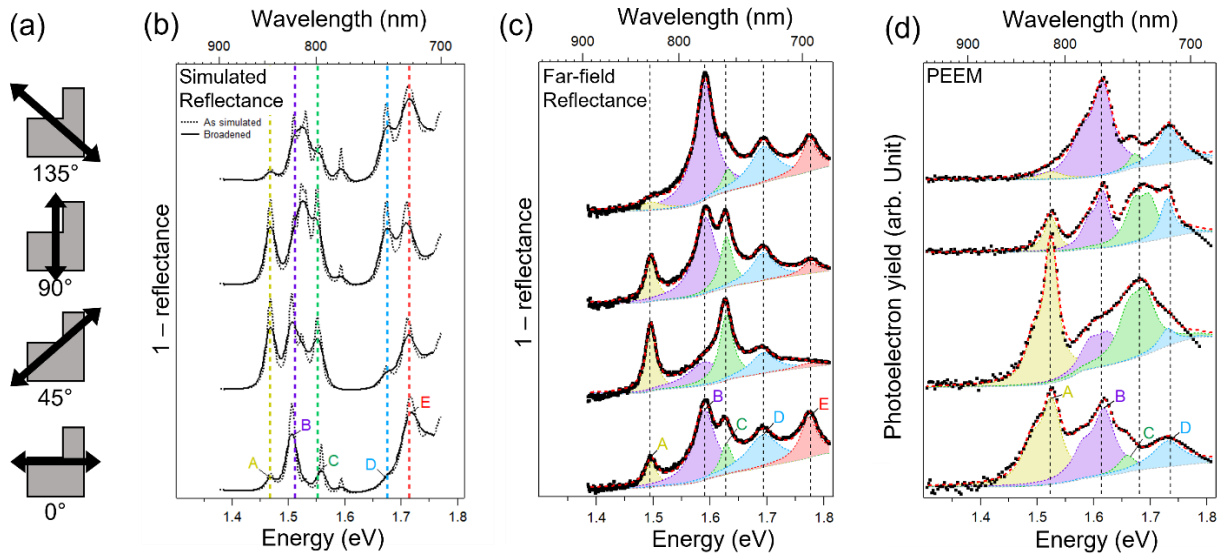

Figure 2: Comparison of metasurface absorption and photoelectron yield spectra at normal incident optical excitation. (a) Schematic of the polarization orientation of the far-field excitation source with respect to the resonator geometry. (b) 1-reflectance spectra of the fabricated Si metasurface calculated using full-wave electromagnetic simulations (COMSOL). Dashed spectra depict the initial simulated spectra while solid spectra are the result of broadening with a Gaussian convolution (FWHM 17.5 meV or ~ 9 nm). Vertical colored dashed lines indicate the general position of each broadened resonance and are labeled A to E to correlate with the far-field reflectance as discussed in the main text. (c) Measured 1-reflectance spectra of the Si metasurface measured with a home-built reflectance set-up. The sample used for this measurement was the same Si metasurface coated with $TiO_2$ and exposed to K during the PEEM measurements. Before the reflectance measurements, the sample was cleaned by annealing in UHV conditions to remove K. (d) Photoelectron yield spectra averaged over many unit-cells measured with PEEM. Spectra in panels (b), (c), and (d) are presented for the four incident polarization orientations depicted in (a).

We next demonstrate real-space imaging of the resonances based on a PEEM imaging scheme. Figure 3a and 3b display large scale photoelectron intensity images containing hundreds of individual resonators located near the center of the fabricated Si metasurface array acquired under optical excitation with deep UV (5.82 eV) and near-IR (1.52 eV) light. Under deep UV laser excitation (Figure 3a), the shapes of resonators are rendered well (compare with SEM image in Figure 1c and resonator design in Figure S2). However, when the excitation is switched to the near-IR, the photoelectron intensity image instead reveals the location of enhanced electric fields for this resonance (A-resonance at 135° polarization orientation). Using the deep UV PEEM image and the measured resonator dimensions from SEM, the borders of a 5x5 subsection of resonators are outlined in white (Figure 3a). Overlaying these resonator positions within the NIR image (Figure 3b) reveals that the enhanced photoemission, and therefore near-field electric fields, are confined within the resonator volume as expected for these semiconducting metasurfaces. This field confinement is further verified through co-illumination experiments provided in Supplementary Figure S4.

We also find significant resonator to resonator variations in the photoelectron intensity image appearing as additional high intensity spots and "inactive" resonators. A few examples are highlighted by the red and yellow circles, respectively, in Figure 3a,b. The shape of the high intensity spots in Figure 3b resembles those of silver nanoparticles imaged with PEEM under resonant conditions in ref. 25 that are uncontrollably agglomerated through the metal deposition process. This observation leads us to speculate that the high intensity resonance spots arise from imperfect side walls of the resonators due to the limited fabrication precision or due to clustered $TiO_2$ on resonator surfaces. As such, PEEM can also serve as a tool to diagnose heterogeneities and defective resonators within a metasurface through direct light matter interactions and improve the engineering aspect of metasurface fabrication.

High-resolution imaging of the photoelectron intensity distributions at different excitation wavelengths demonstrates the sensitivity of this approach to the near-field electric field profiles of the different modes. Figure 3c-f display averaged photoelectron intensity images for a single resonator unit under excitation at the four resonance wavelengths identified in Figure 2d for a 135° polarization orientation. These images are created by superimposing 25 resonators within a single image to account for fabrication variations and improve statistics. Here, we observe that the near-field photoelectron intensity distribution is sensitive to excitation energy. For the A- and B-resonances, the near-field distributions are highly symmetric as expected for the out-of-plane magnetic and electronic dipole modes, respectively. Meanwhile, the C- and D-resonances display non-symmetric and highly structured near-field distributions, as expected given the higher order of the modes that make up these resonances. In addition to excitation wavelength, photoelectron intensity images also capture the polarization dependence of the optical modes and match well with full wave simulations, which will be discussed in the following paragraphs.

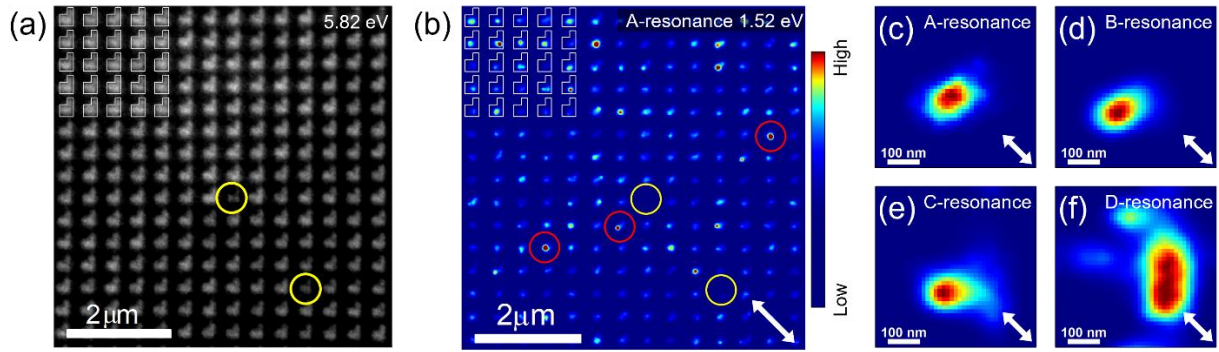

Figure 3: Photoelectron intensity confinement via energy dependent nanoscale mode imaging. (a,b) Large field of view photoelectron intensity images of the metasurface array imaged using deep ultraviolet (5.82 eV, a) and A-resonance (1.52 eV, b) excitation energies. The boundaries for a 5x5 subset of the array are outlined in white in (a) and transposed to (b) to show the confinement of photoelectron intensity within the resonator boundaries. Yellow circles in (a) identify defective resonators from fabrication, which are found in (b) to be inactive. Abnormal high intensity spots in (b) are identified with red circles that may arise from additional fabrication defects or particle formation. (c), (d), (e), (f) Resonator averaged photoelectron intensity images at resonant excitation energies identified for 135° polarization orientation, 1.52 eV (c), 1.62 eV (d), 1.69 eV (e), and 1.73 eV (f). Polarization orientation is indicated by white double arrows in (b), (c), (d), (e) and (f).

Enabled by the far-field optical excitation, polarization resolved near-field photoelectron images are acquired for the resonances supported by this Si metasurface and found to match well with simulated profiles of the square of the electric field intensity ($|E|^4$). Near-field photoelectron intensity images obtained at the four resonant wavelengths (A-D) for a 0° polarization orientation are shown in Figure 4a-d averaged over at least 20 individual resonators. Additionally, supplementary video S1 displays the complete PEEM spectral hypercube for a 0° polarization. Here, each resonant wavelength generates a distinct spatial distribution with an enhancement in the photoelectron intensity of 3-5 times higher than the area outside the resonator boundaries, demonstrating the near-field enhancement of two-photon photoemission that arises from the optical resonances. Importantly, these spatial distributions are visually dissimilar from those obtained at a 135° polarization orientation (Figure 3 c-f) exemplifying the capability to achieve polarization resolved near-field imaging through far-field excitation.

Moreover, at a 0° polarization orientation, each resonance observed experimentally for the fabricated Si metasurface can be mapped to a singular mode from simulation (Figure 2). As a result, the photoelectron intensity distributions in Figure 4a-d are found to recreate the simulated profiles of the square of the electric field intensity for each of these individual modes exceptionally well (Figure 4e-h). The simulated field profiles in Figure 4e-h are 2D planes extracted from the middle height of a resonator in an infinite array under excitation at a 0° polarization orientation, regions outside the resonator boundaries are set to zero. Photoelectron intensity images for the quasi-BIC A- and B-resonances (Figure 4a and b) exhibit spherical distributions that closely resemble the fields expected from simulation (Figure 4e and f), which places these enhancements near the center of the resonator unit. Extending to the higher order modes (C- and D-resonances, Figure 4c and d) we observe an extension and then splitting of the photoelectron intensity towards the upper right portion of the resonator, a trend that is well captured in the corresponding simulated field distributions (Figure 4g and h). Generally, the photoelectron intensity distributions appear spatially broadened relative to the simulated field enhancements, and some portions of fine structure are not captured, such as the splitting into three high intensity regions for the highest energy D-mode. Such discrepancies are likely due to imperfections in the fabricated resonator geometries and array spacing, in addition to the broader linewidth of the excitation source in experiment. Nonetheless, we capture high resolution near-field images through photoelectron imaging that are sensitive to wavelength and are well correlated with the expected electric field enhancements from full wave simulations.

Near-field photoelectron intensity imaging at a 45° polarization orientation for the same four resonances (A-D) reveals further distinct distributions that well recreate the expected field enhancements, despite the overlapping modes of some of the resonances. As seen in Figure 4i and 4m, the photoelectron intensity distribution at the lowest energy A-resonance is spherical in nature and located near the center of the resonator unit. However, when increasing the excitation energy, there is a clear elongation and splitting of the photoelectron intensity along the polarization axis (Figure 4j-l). This evolution in field distribution is consistent with the simulated fields for the modes corresponding to these resonances (Figure 4m-p). We note that the B- and C- resonances at a 45° polarization orientation (Figure 4j and k) consist of two overlapping modes (Figure 2). The simulated field distributions depicted in Figure 4n and o for these resonances are the single mode profiles that best match each photoelectron image, although the latter is likely a mixture of the two overlapping modes. Photoelectron images and the corresponding simulated profiles of the square of the electric field intensity for all resonance and modes can be found in Supplementary Figures S5, S6, S7, and S8 sorted by incident polarization orientation. Comparison of the photoelectron images and simulated field profiles at 90° and 135° polarization angles (Figure S7 and S8) reveals poorer matching than for 0° and 45° (Figure 4, S5, and S6) due to an increase in number of excited modes. In the former, the poorer matching results from the fact that the field distribution will depend on the coherent superposition resulting from interference between these simultaneously excited modes. The overall fields will therefore depend on the relative amplitudes, *Q*-factors, and relative phases of the modes being excited, which is extremely difficult to predict using simulations. Altogether, we demonstrate high resolution polarization resolved near-field imaging through photoelectron microscopy and find reasonable agreement with the electric field enhancements from full wave simulations.

Photoemission processes are typically regarded as surface sensitive since the inelastic mean free path (IMFP) of photoelectrons in most materials is expected to be on the order of a few nanometers.[54,55] Though, in the case of low energy electrons (< 10 eV), as in this work, IMFPs of 10-100 nm have been obtained experimentally for Au and $Al_2O_3$, and predicted from theory for many more elements.[56,57,58,59] Nevertheless, the number of escaping electrons is expected to decay exponentially as a function of depth and we thus expect the PEEM signal to originate primarily from the $TiO_2$ layer or sub-monolayer K surface layer. Even so, the photoelectron intensity images presented in Figure 4 correlate well with simulated profiles of the square of the electric field intensity extracted as 2D cross-sections at the middle height of a Si resonator. A degree of sensitivity along this z direction may arise from photoelectron emission from the sidewalls of the resonators, as these electrons could escape to vacuum without traversing the full resonator height. A more probable interpretation is that the field enhancements within the resonator can excite photoelectrons from the surface layers, thereby eliciting photoelectron contrasts from light-matter interactions occurring at depths well beneath where the photoelectrons themselves are emitted. Previous works have demonstrated this depth sensitivity beyond typical photoelectron escape lengths arising from similar phenomena in buried dielectric interfaces.[60,61] Finally, we note that the nanostructures in this study are ill-suited for detailed examination of z sensitivity given that the resonator height is on the order of half the excitation wavelength and that future studies on taller nanostructures (height > λ) designed for more structured variation of the fields along the z axis would be more appropriate.

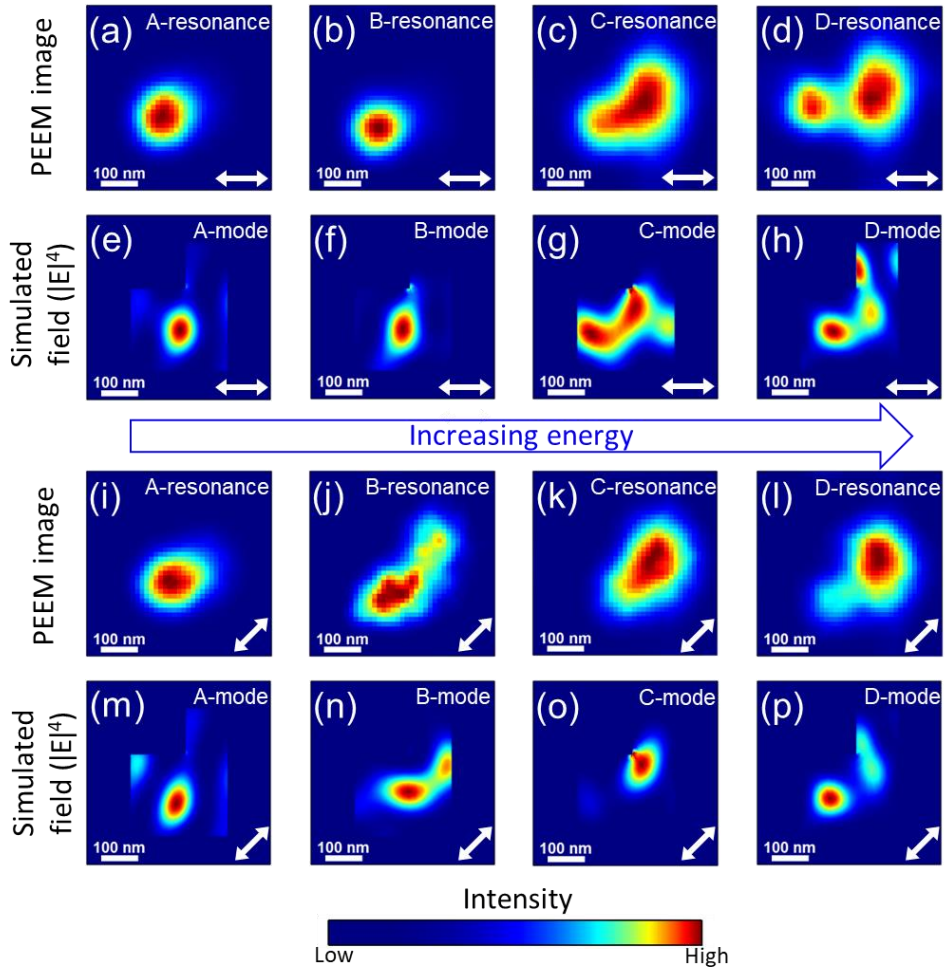

Figure 4: Excitation energy dependent and polarization resolved nanoscale mode imaging and simulated electromagnetic field profiles of resonators. (a), (b), (c), (d) Resonator averaged photoelectron intensity images at resonant excitation energies for a 0° polarization orientation, 1.52 eV (a), 1.62 eV (b), 1.67 eV (c), and 1.73 eV (d). (e), (f), (g), (h) Simulated profiles of the square of the electric field intensity ($|E|^4$) for the corresponding modes at a 0° polarization orientation. (i), (j), (k), (l) Resonator averaged photoelectron intensity images at resonant excitation energies for a 45° polarization orientation, 1.53 eV (i), 1.61 eV (j), 1.67 eV (k), and 1.75 eV (l). (m), (n), (o), (p) Simulated profiles of the square of the electric field intensity ($|E|^4$) for the corresponding modes at a 45° polarization orientation. Polarization orientations are also indicated by the white double arrows.

Thus far all photoelectron spectra and images have originated from regions within the resonator array distant from array edges as the photonic modes studied in this work are collective modes (*i.e.*, involve neighboring resonators). However, we anticipate that at the boundaries of the array, the confinement of the electromagnetic field will vary as a function of distance from the edge until the resonances are fully established.[15] To examine this collective nature, we investigate the evolution of the photoelectron yield spectra over the first few resonators near the array edge.

Within this regime, we observe an enhancement of the quasi-BIC containing B-resonance in the outermost resonators followed by a polarization dependent quenching of the second column/row of the array. Figure 5a displays a photoelectron image collected at the upper-left corner of a large resonator array under excitation at the B-resonance wavelength with a 90° polarization angle [Supplementary Figure S9 shows the same region imaged under DUV (5.82 eV) excitation]. Resonators making up the outermost border of the array (R1 and C1) exhibit significantly enhanced photoelectron intensity relative to their interior counterparts. We interpret this enhancement as a weakening of the quasi-BIC confinement due to lack of neighboring elements resulting in better in-coupling of light to these edge resonators. Moreover, we note a strong suppression of the B-resonance photoelectron intensity in the second column (C2) that is quickly recovered in the third column (C3). This suppression is not a consequence of defective fabrication, but rather a unique interaction occurring within the near edge region of the resonator array parallel to the incident polarization. For when the incident light polarization is rotated parallel to the top edge of the array (0° polarization), the resonators in C2 exhibit comparable photoelectron intensity to their interior counterparts, and instead the resonators in R2 become suppressed, as shown in Figure S10a. The enhancement, suppression, and then recovery of the B-resonance photoelectron intensity across the near edge of the array (C1-3) can also be visualized by plotting the photoelectron yield spectra as a function of column number (distance from left edge) as seen in Figure 5b. Through fits to these spectra, we ascertain that while the described near edge evolution is unique to the B-resonance the remaining resonances do undergo more gradual changes, with all trending towards bulk as the distance from the array edge increases.

From the evolution of the photoelectron yield spectra and relative peak areas across this range, we conclude that approximately eight resonators, in both propagation directions, are necessary to fully establish the collective array modes, corroborating the result of previous works. [15,20] This is observed in Figure 5b when the photoelectron yield spectra (and corresponding fits) continue to evolve beyond C3 before converging after about eight resonators (C8) where they closely resemble the bulk spectra (Figure 2d). This trend is more clearly visualized when the relative peak areas for each resonance are plotted as a function of column number (Figure 5c). Here, the dramatic changes across the near edge region (C1-3) quickly give way to more consistent trends spanning C4-7 (termed the 'transition' region) before stabilizing at C8 and beyond (termed the 'bulk' region) at comparable values to the bulk shown in Figure 2d. Propagating away from the top array edge (*i.e.*, row number) in Figure 5d demonstrates the same general trends, that is, significant variation across the near edge region, more gradual changes within the transition region, and stabilization after eight resonators. Evolution of the photoelectron yield spectra and relative peak areas for the same region under illumination at a 0° polarization orientation also demonstrate the trends noted here (Figure S10). This data clearly supports the notion that the excited photonic modes of the resonators near the array edges are different from the photonic modes in the bulk of the metasurface (*i.e.*, the center of the array). By capturing photoelectron yield spectra at each point of the image, PEEM enables us to draw such conclusions and study how these collective optical resonances form within the metasurfaces under far-field optical excitations.

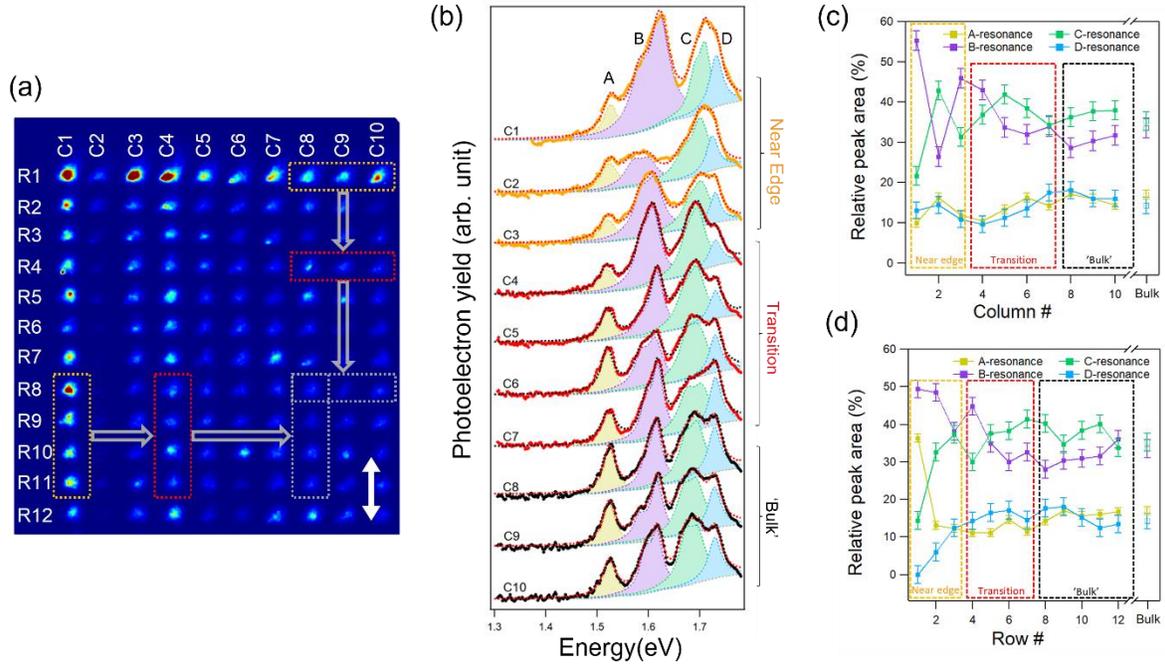

Figure 5: Evolution of the coherent interactions from the edges to the middle of the resonator array. (a) Photoelectron image of the upper-left corner of a large metasurface array acquired at 1.62 eV excitation (B-resonance) and a 90° polarization orientation (indicated by white double arrow). The rows and columns are designated as R1-R12 and C1-C10, respectively, within the image. Supplementary Figure S9 shows the same region imaged under DUV (5.82 eV) excitation for better viewing of the array corner. (b) The area averaged spectra of the resonators as a function of the distance from the left array edge (column number). The spectra are averaged for 4 unit-cells along the vertical direction beginning at R8, as indicated by the yellow, red, and grey doted rectangles. (c, d) The relative peak area for each resonance as a function of the column (c) or row (d) number (*i.e.*, distance from left or top array edge, respectively). The relative peak areas in (d) come from fits to the spectra averaged over 3 unit-cells along the horizontal direction beginning at C8, as indicated in (a). Relative peak areas from bulk come from the spectral fits in Figure 2d. Error bars correspond to the standard deviation observed for each resonance across the Bulk region averaged for Figure 2d.

In summary, we have presented polarization resolved concurrent spectroscopy and high-resolution imaging of the near-field distribution of photonic resonances in a Si metasurface using photoelectron emission microscopy (PEEM). By capitalizing on far-field optical excitation we gain insights into the anisotropic nature of light-matter interactions in semiconductor nanostructures in the form of polarization dependent collective modes in broken-symmetry Si metasurfaces. Upon exciting and imaging many resonators simultaneously we uncover an unanticipated polarization dependent field enhancement switching near the edges of a larger metasurface array and that coupling between eight resonators is needed to establish quasi-BIC resonances in this Si metasurface. In expanding the application space of PEEM beyond plasmonic nanostructures, we establish a powerful route for the metrology of local and non-local photonic resonances in semiconductor nanostructures.

**ACKNOWLEDGEMENT:**

We thank P. Mantos, A. Jarzembski, and G. Copeland for their measurement support and A. Cerjan for discussions. A.B., T.E.B., and T.O. acknowledge support from the Laboratory Directed Research and Development program at Sandia National Laboratories. S.D.G., C.F.D., M.B.S., I.B., and R.S. acknowledge support from the U.S. Department of Energy, Office of Basic Energy Sciences, Division of Materials Sciences and Engineering. This work was performed in part at the Center for Integrated Nanotechnologies, an Office of Science User Facility operated for the US Department of Energy (DOE) Office of Science. Sandia National Laboratories is a multimission laboratory managed and operated by National Technology and Engineering Solutions of Sandia, LLC., a wholly owned subsidiary of Honeywell International, Inc., for the U.S. Department of Energy's National Nuclear Security Administration under contract DE-NA0003525. This paper describes objective technical results and analysis. Any subjective views or opinions that might be expressed in the paper do not necessarily represent the views of the U.S. Department of Energy or the United States Government.

This article has been authored by an employee of National Technology & Engineering Solutions of Sandia, LLC under Contract No. DE-NA0003525 with the U.S. Department of Energy (DOE). The employee owns all right, title and interest in and to the article and is solely responsible for its contents. The United States Government retains and the publisher, by accepting the article for publication, acknowledges that the United States Government retains a non-exclusive, paid-up, irrevocable, world-wide license to publish or reproduce the published form of this article or allow others to do so, for United States Government purposes. The DOE will provide public access to these results of federally sponsored research in accordance with the DOE Public Access Plan https://www.energy.gov/downloads/doe-public-access-plan.

**REFERENCES:**

[1] Novotny, L.; Sánchez, E. J.; & Sunney Xie, X., Near-field optical imaging using metal tips illuminated by higher-order Hermite–Gaussian beams. Ultramicroscopy, 1998, 71(1), 21–29.

[2] Mühlschlegel, P.; Eisler, H.-J.; Martin, O. J. F.; Hecht, B.; & Pohl, D. W., Resonant Optical Antennas. Science, 2005, 308(5728), 1607–1609.

[3] Bharadwaj, P.; Deutsch, B.; & Novotny, L., Optical Antennas. Advances in Optics and Photonics, 2009, 1(3), 438–483. 8

[4] Holloway, C. L.; Kuester, E. F.; Gordon, J. A.; O'Hara, J.; Booth, J.; & Smith, D. R., An Overview of the Theory and Applications of Metasurfaces: The Two-Dimensional Equivalents of Metamaterials. IEEE Antennas and Propagation Magazine, 2012, 54(2), 10–35.

[5] Ginn, J. C.; Brener, I.; Peters, D. W.; Wendt, J. R.; Stevens, J. O.; Hines, P. F.; Basilio, L. I.; Warne, L. K.; Ihlefeld, J. F.; Clem, P. G.; & Sinclair, M. B., Realizing Optical Magnetism from Dielectric Metamaterials. Physical Review Letters, 2012, 108(9), 97402.

[6] Brener, I.; Liu, S.; Staude, I.; Valentine, J.; Holloway, C., Dielectric Metamaterials: Fundamentals, Designs and Applications; Woodhead Publishing, (2019).

[7] Yablonovitch, E., Inhibited Spontaneous Emission in Solid-State Physics and Electronics, Phys. Rev. Lett. 1987, 58, 2059-2062.

[8] Yablonovitch, E.; Gmitter, T. J.; & Leung, K. M., Photonic band structure: The face-centered-cubic case employing nonspherical atoms. Physical Review Letters, 1991, 67(17), 2295–2298.

[9] Chen, X.; Zhang, Y.; Cai, G.; Zhuo, J.; Lai, K.; Ye, L., All-dielectric metasurfaces with high Q-factor Fano resonances enabling multi-scenario sensing, Nanophotonics, 2022, 11(20), 4537-4549.

[10] Kamali, S. M.; Arbabia, E.; Arbabi, A.; Faraon, A., A review of dielectric optical metasurfaces for wavefront control; Nanophotonics 2018, 7(6), 1041-1068.

[11] Genevet, P.; Capasso, F., Holographic optical metasurfaces: a review of current progress Rep. Prog. Phys. 2015, 78, 024401.

[12] Krasnok, A.; Tymchenko, M.; Alu, A., Nonlinear Metasurfaces: A Paradigm Shift in Nonlinear Optics. Materials Today 2018, 21(1), 8-21.

[13] Solntsev, A. S., Agarwal, G. S. & Kivshar, Y. S., Metasurfaces for quantum photonics. Nat. Photonics, 2021, 15, 327–336.

[14] Santiago-Cruz, T.; Gennaro, S. D.; Mitrofanov, O.; Addamane, S.; Reno, J.; Brener, I.; Chekhova, M. V., Resonant metasurfaces for generating complex quantum states, Science, 2022, 377, 6609, 991-995.

[15] Campione, S.; Liu, S.; Basilio, L.I.; Warne, L.K.; Langston, W.L.; Luk, T.S.; Wendt, J.R.; Reno, J.L.; Keeler, G.A.; Brener, I. and Sinclair, M.B., Broken symmetry dielectric resonators for high quality factor Fano metasurfaces. ACS Photonics, 2016, 3(12), 2362-2367.

[16] Liu, Z.; Xu, Y.; Lin, Y.; Xiang, J.; Feng, T.; Cao, Q.; Li, J.; Lan, S.; & Liu, J., High-Q Quasibound States in the Continuum for Nonlinear Metasurfaces, Physical Review Letters, 2019, 123(25), 253901.

[17] García de Abajo, F. J., Optical excitations in electron microscopy. Reviews of Modern Physics, 2010, 82(1), 209–275.

[18] Coenen, T.; van de Groep, J.; Polman, A., Resonant Modes of Single Silicon Nanocavities Excited by Electron Irradiation, ACS Nano 2013, 7, 1689–1698.

[19] Kociak, M.; Sté´phan, O., Mapping plasmons at the nanometer scale in an electron microscope, Chem. Soc. Rev., 2014, 43, 3865-3883.

[20] Dong, Z.; Mahfoud, Z.; Paniagua-Domínguez, R.; Wang, H.; Fernández-Domínguez, A. I.; Gorelik, S.; Ha, S. T.; Tjiptoharsono, F.; Kuznetsov, A. I.; Bosman, M.; Yang, J. K. W., Nanoscale mapping of optically inaccessible bound-states-in-the-continuum, Light: Science & Applications, 2022, 11, 20.

[21] Polman, A.; Kociak, M.; García de Abajo, F. J., Electron-beam spectroscopy for nanophotonics, Nature Mater, 2019, 1158, 1158–1171.

[22] Habteyes, T. G., Staude, I., Chong, K. E., Dominguez, J., Decker, M., Miroshnichenko, A., Kivshar, Y. and Brener, I., Near-field mapping of optical modes on all-dielectric silicon nanodisks. ACS Photonics, 2014, 1(9), pp.794-798.

[23] Tamagnone, M.; Ambrosio, A.; Chaudhary, K.; Jauregui, L. A.; Kim, P.; Wilson, W. L.; Capasso, F., Ultra-confined mid-infrared resonant phonon polaritons in van der Waals nanostructures, Sci. Adv. 2018, 4, eaat7189.

[24] Yin, L.; Vlasko-Vlasov, V. K.; Rydh, A.; Pearson, J.; Welp, U.; Chang, S.-H.; Gray, S. K.; Schatz, G. C.; Brown, D. B.; Kimball, C. W., Surface plasmons at single nanoholes in Films, Appl. Phys. Lett. 2004, 85, 467.

[25] Kubo, A.; Onda, K.; Petek, H.; Sun, Z.; Jung, Y. S.; Kim, H. K., Femtosecond Imaging of Surface Plasmon Dynamics in a Nanostructured Silver Film, Nano Lett. 2005, 5, 6, 1123–1127.

[26] Cinchetti, M.; Gloskovskii, A.; Nepjiko, S. A.; Schönhense, G.; Rochholz, H.; Kreiter, M., Photoemission Electron Microscopy as a Tool for the Investigation of Optical Near Fields, Phys. Rev. Lett., 2005, 95, 047601.

[27] Frank, B., Kahl, P., Podbiel, D., Spektor, G., Orenstein, M., Fu, L., Weiss, T., Horn-von Hoegen, M., Davis, T. J., Meyer zu Heringdorf, F.-J., & Giessen, H., Short-range surface plasmonics: Localized electron emission dynamics from a 60-nm spot on an atomically flat single-crystalline gold surface. Science Advances, 2017, 3(7), e1700721.

[28] Sun, Q., Zu, S., & Misawa, H., Ultrafast photoemission electron microscopy: Capability and potential in probing plasmonic nanostructures from multiple domains, The Journal of Chemical Physics, 2020, 153(12), 120902.

[29] Zhang, L.; Kubo, A.; Wang, L.; Petek, H.; Seideman, T., Imaging of surface plasmon polariton fields excited at a nanometer-scale slit, Phys. Rev. B, 2011, 84, 245442.

[30] Kahl, P.; Wall, S.; Witt, C.; Schneider, C.; Bayer, D.; Fischer, A.; Melchior, P.; Horn-von Hoegen, M.; Aeschlimann, M.; Meyer zu Heringdorf, F. J., Normal-Incidence Photoemission Electron Microscopy (NI-PEEM) for Imaging Surface Plasmon Polaritons, Plasmonics, 2014, 9, 1401–1407.

[31] Gong, Y.; Joly, A. G.; El-Khoury, P. Z.; Hess, W. P., Nonlinear Photoemission Electron Micrographs of Plasmonic Nanoholes in Gold Thin Films, J. Phys. Chem. C 2014, 118, 25671-25676.

[32] Fitzgerald, J. P. S.; Word, R. C.; Saliba, S. D.; Könenkamp, R., Photonic near-field imaging in multiphoton photoemission electron microscopy, Phys. Rev. B, 2013, 87, 205419.

[33] Hartmann, H.; Barke, I.; Friedrich, A.; Plötz, P.-A.; Bokareva, O. S.; Bahrami, M.; Oldenburg, K.; Elemans, J. A. A. W.; Irsig, R.; Meiwes-Broer, K.-H.; Kühn, O.; Lochbrunner, S.; Speller, S., Mapping Long-Lived Dark States in Copper Porphyrin Nanostructures, J. Phys. Chem. C, 2016, 120, 30, 16977–16984.

[34] Stallberg, K.; Lilienkamp, G.; Daum, W., Plasmon–Exciton Coupling at Individual Porphyrin-Covered Silver Clusters, J. Phys. Chem. C, 2017, 121, 25, 13833–13839.

[35] Aeschlimann, M.; Brixner, T.; Differt, D.; Heinzmann, U.; Hensen, M.; Kramer, C.; Lükermann, F.; Melchior, P.; Pfeiffer, W.; Piecuch, M.; Schneider, C.; Stiebig, H.; Strüber, C.; Thielen, P., Perfect absorption in nanotextured thin films via Anderson-localized photon modes. Nature Photonics, 2015, 9(10).

[36] Mårsell, E.; Boström, E.; Harth, A.; Losquin, A.; Guo, C.; Cheng, Y.-C.; Lorek, E.; Lehmann, S.; Nylund, G.; Stankovski, M.; Arnold, C. L.; Miranda, M.; Dick, K. A.; Mauritsson, J.; Verdozzi, C.; L'Huillier, A.;  Mikkelsen, A., Spatial Control of Multiphoton Electron Excitations in InAs Nanowires by Varying Crystal Phase and Light Polarization. Nano Letters, 2018, 18(2), 907–915.

[37] Xu, C.; Yong, H. W.; He, J.; Long, R.; Cadore, A. R.; Paradisanos, I.; Ott, A. K.; Soavi, G.; Tongay, S.; Cerullo, G.; Ferrari, A. C.; Prezhdo, O. V.; Loh, Z.-H., Weak Distance Dependence of Hot-Electron-Transfer Rates at the Interface between Monolayer MoS2 and Gold. ACS Nano, 2021, 15(1), 819–828.

[38] Hsu, C. W.; Zhen, B.; Stone, A. D.; Joannopoulos, J. D.; Soljac¡ic, M., Bound states in the continuum, Nature Reviews Materials, 2016, 1, 16048.

[39] Koshelev, K.; Lepeshov, S.; Liu, M.; Bogdanov, A.; Kivshar, Y., Asymmetric Metasurfaces with High-Q Resonances Governed by Bound States in the Continuum, Phys. Rev. Lett., 2018, 121, 193903.

[40] Koshelev, K.; Bogdanov, A.; Kivshar, Y., Engineering with Bound States in the Continuum, Optics and Photonics News, 2020, 31, 38-45.

[41] Hüfner, S., Photoelectron Spectroscopy - Principles and Applications, Springer-Verlag Berlin Heidelberg, 2003.

[42] Melchior, P.; Bayer, D.; Schneider, C.; Fischer, A.; Rohmer, M.; Pfeiffer, W.; & Aeschlimann, M., Optical near-field interference in the excitation of a bowtie nanoantenna. 2011, Physical Review B, 83(23), 235407.

[43] Yu, H.; Sun, Q.; Ueno, K.; Oshikiri, T.; Kubo, A.; Matsuo, Y.; Misawa, H., Exploring Coupled Plasmonic Nanostructures in the Near Field by Photoemission Electron Microscopy, ACS Nano 2016, 10, 11, 10373–10381.

[44] Bauer, E., Surface Microscopy with Low Energy Electrons, Springer New York, 2014.

[45] Genco, A.; Cruciano, C.; Corti, M.; McGhee, K. E.; Ardini, B.; Sortino, L.; Huttenhofer, L.; Virgili, T.; Lidzey, D. G.; Maier, S. A.; Bassi, A.; Valentini, G.; Cerullo, G.; Manzoni, C., k-Space Hyperspectral Imaging by a Birefringent Common-Path Interferometer, ACS Photonics 2022, 9, 3563-3572.

[46] Staude, I.; Schilling, J., Metamaterial-inspired silicon nanophotonics, Nature Photonics, 2017, 11, 274–284.

[47] Douillard, L., & Charra, F., Photoemission electron microscopy, a tool for plasmonics. Journal of Electron Spectroscopy and Related Phenomena, ,2013, 189, 24–29.

[48] Plotnik, Y.; Peleg, O.; Dreisow, F.; Heinrich, M.; Nolte, S.; Szameit, A.; Segev, M., Experimental Observation of Optical Bound States in the Continuum, Phys. Rev. Lett., 2011, 107, 183901.

[49] Cerjan, A.; Jörg, C.; Vaidya, S.; Augustine, S.; Benalcazar, W. A.; Hsu, C. W.; von Freymann, G.; Rechtsman, M. C., Observation of bound states in the continuum embedded in symmetry bandgaps, Sci. Adv., 2021, 7, eabk1117.

[50] Liu, S.; Vaskin, A.; Addamane, S.; Leung, B.; Tsai, M. C.; Yang, Y.; Vabishchevich, P. P.; Keeler, G. A.; Wang, G.; He, X.; Kim, Y.; Hartmann, N. F.; Htoon, H.; Doorn, S. K.; Zilk, M.; Pertsch, T.; Balakrishnan, G.; Sinclair, M. B.; Staude, I.; Brener, I., Light-Emitting Metasurfaces: Simultaneous Control of Spontaneous Emission and Far-Field Radiation, Nano Lett., 2018, 18, 11, 6906–6914.

[51] Overvig, A. C.; Malek, S. C.; Carter, M. J.; Shrestha, S.; Yu, N., Selection rules for quasibound states in the continuum, Phys. Rev. B, 2020, 102, 035434.

[52] Bostwick, A.; Ohta, T.; Seyller, T.; Horn, K.; Rotenberg, E., Quasiparticle dynamics in graphene, Nature Physics, 2007, 3, 36-40.

[53] Sarma, R.; Campione, S.; Goldflam, M.; Shank, J.; Noh, J.; Smith, S.; Ye, P. D.; Sinclair, M.; Klem, J.; Wendt, J.; Ruiz, I.; Howell, S. W.; Brener, I., Low dissipation spectral filtering using a field-effect tunable III–V hybrid Metasurface, App. Phy. Lett, 2018, 113, 061108.

[54] Tanuma, S.; Powell, C. J.; Penn, D. R., Calculations of electron inelastic mean free paths. II. Data for 27 elements over the 50–2000 eV range, Surface & Interface Analysis, 1991, 17, 13, 911-926.

[55] Kuhr, J. C.; Fitting, H. J., Monte Carlo simulation of electron emission from solids, Journal of Electron Spectroscopy and Related Phenomena, 1999, 105, 2–3, 257-273.

[56] Ding, Z.-J.; Shimizu, R., Inelastic collisions of kV electrons in solids. Surface Science, 1989, 222(2), 313–331.

[57] W. Pong, Photoemission from Al–Al2O3 Films in the Vacuum Ultraviolet Region. J. Appl. Phys., 1969, 40 (4): 1733–1739.

[58] Sze, S. M.; Moll, J. L.; Sugano, T., Range-energy relation of hot electrons in gold. Solid-State Electronics, 1964, 7(7), 509–523.

[59] Ridzel, O. Yu.; Astašauskas, V.; Werner, W. S. M., Low energy (1–100 eV) electron inelastic mean free path (IMFP) values determined from analysis of secondary electron yields (SEY) in the incident energy range of 0.1–10 keV. Journal of Electron Spectroscopy and Related Phenomena, 2020, 241, 146824.

[60] Berg, M.; Liu, F.; Smith, S.; Copeland, R. G.; Chan, C. K.; Mohite, A. D.; Beechem, T. E.; Ohta, T., Imaging atomically thin semiconductors beneath dielectrics via deep ultraviolet photoemission electron microscopy, Phys. Rev. Applied, 2019, 12, 064064.

[61] Beechem, T. E.; Smith, S. W.; Copeland, R. G.; Liu, F.; Ohta, T., Spectral and Polarization Based Imaging in Deep-Ultraviolet Excited Photoelectron Microscopy, Review of Scientific Instruments, 2022, 93, 053701.

# Supporting information: Near-field Imaging of Optical Resonances in Si Metasurfaces Using Photoelectron Microscopy

Alex Boehm[1], Sylvain D. Gennaro[1,2], Chloe F. Doiron[1,2], Thomas E. Beechem[3], Michael B. Sinclair[1], Igal Brener[1,2], Raktim Sarma[1,2], Taisuke Ohta[1]*

1 Sandia National Laboratories, Albuquerque, NM 87185

2 Center for Integrated Nanotechnologies, Albuquerque, NM 87185

3 School of Mechanical Engineering and Birck Nanotechnology Center, Purdue University, West Lafayette, IN 47907

**METHODS:**

**Numerical simulation**

Full-wave 3D electromagnetic simulations of the Si metasurface were performed using finite element method (COMSOL with the Electromagnetic Waves, Frequency Domain interface). Each model simulates an infinite (using periodic boundary conditions) square lattice of silicon resonators with a width of 265 nm and depth of 225 nm with a notch width and depth of 115 nm and 125 nm, respectively, and resonator height of 300 nm (see Supplementary Figure S2). The lattice period is 413 nm. The structure is illuminated with a plane wave at normal incidence. For these simulations, the silicon was treated as non-hydrogenated amorphous silicon[1], while the Au layer used a Drude model ($\varepsilon = 1 - \frac{\omega_p^2}{\omega^2 - i\omega\Gamma}$) with a plasma frequency ($\omega_p$) of 72,800 cm$^{-1}$ and damping parameter ($\Gamma$) of 215 cm$^{-1}$. The SiO$_2$ layers used a refractive index of 1.44 and thickness of 30 nm. The Al$_3$O$_3$ layers used a refractive index of 1.76 and thicknesses of 15 nm. From these simulations, we extracted horizontal electric field cross sections at 150 nm away from the top of the resonator. For these plots, the field outside the resonator was ignored, since there is no material there for photoelectron emission.

**Sample fabrication**

The Si metasurface were fabricated using electron-beam lithography of polycrystalline Si films (300 nm thick), sputtered on a stack of Al$_2$O$_3$ (15nm), SiO$_2$ (30nm), Al$_2$O$_3$ (15nm) and Au (110nm) thin films, deposited on a Si wafer. The surface of the Si metasurface was then conformally coated with a 10nm-thick TiO$_2$ layer using atomic layer deposition (ALD). Both Au and TiO$_2$ were introduced to avoid sample charging during PEEM imaging due to their higher electric conductivities; we observed that metasurfaces fabricated on traditional Silicon-on-insulator substrate led to sample charging during PEEM imaging, which deteriorated the image quality.

We also verified that the deposition of the TiO$_2$ layer has minimal impact on the optical resonance properties of the Si metasurface based on reflectance measurement taken before and after the TiO$_2$ deposition. On some occasions, the resonance features of the Si metasurface sharpened after the TiO$_2$ deposition presumably due to the sample being heated during the ALD process, which may have improved the crystallinity of the sputtered Si.

**Far-field reflectivity**

We measured the linear spectra of the semiconductor metasurface using a homebuilt reflectance system. The sample used for this measurement was the same Si metasurface coated with $TiO_2$ and exposed to K during the PEEM measurements. However, before the far-field reflectance measurements were acquired, the sample was cleaned by annealing in UHV conditions. A polarized broadband white light emission from a thermal source is focused onto the sample using an achromatic lens of focal length 50 mm. The scattered light, collected from the same lens, is then rerouted toward the entrance slit of a spectrometer. The low numerical aperture (0.2) of our imaging system limits the excitation angle of incident light, which preserves the quality factor of the quasi-BIC modes.

**PEEM imaging**

We conducted near-field imaging using a LEEM-III system (Elmitec Elektronenmikroskopie GmbH), operated in the photoelectron emission microscopy (PEEM) mode. Our PEEM instrument is connected to the output of a tunable near-infrared (IR) Ti:Sapphire laser oscillator (normal incident, ~100 femtosecond pulse, Coherent Inc.), a deep ultraviolet (UV) laser (normal incident, 213 nm continuous wave, Toptica Photonics), and a deep UV-visible incoherent light source (incident angle of 73° relative to the sample surface normal, continuous wave, Energetiq Technology) with a Czerny–Turner monochromator (Acton Research Corporation). The UV and near-IR lasers are focused onto the sample surface (to spot sizes of 50 µm to 100 µm, respectively) using a fused silica lens of 750 mm focal length. The polarizations of the lasers are rotated with half waveplates. More details on the deep UV laser and the incoherent light sources can be found in Berg, *et al.*[2] and Sharma, *et al*.[3]

The photoelectron yield spectra shown in Figure 2 are captured by changing the fundamental wavelength of the near-IR laser, and sequentially recording the photoelectron images, while maintaining the laser power constant for all wavelengths (typically 50 mW or less). The photoelectron yield spectra are then extracted from the specific pixel, or the pixels averaged in the data cube (illustrated in the lower left part of Fig. 1). We limit the laser power to be no more than 100 mW below which minimal space charge image degradation is found.[4]

Prior to a PEEM measurement, the Si metasurface is loaded inside the PEEM instrument in an ultrahigh vacuum (base pressure $3 \times 10^{-11}$ Torr in LEEM-III system), and then annealed overnight at ~150°C to remove water and other chemical species physisorbed from the air. Without unloading the sample, we then deposit no more than one atomic layer of potassium (K), provided by the alkali dispenser (SAES Getters S.p.A.) attached to the PEEM instrument while the pressure not exceeding $1\text{-}2 \times 10^{-10}$ Torr. Introducing K lowers the work function of the metasurface to ~2.7 eV. This value was determined via two independent measurements from the photoelectron onset using the deep UV incoherent light source (one-photon photoemission) and using the near-IR Ti:Sapphire laser (two-photon photoemission).

The ultrahigh vacuum ensures that the oxidation of K proceeds slowly such that the quality of K is preserved during the duration of the PEEM measurement, which is typically 2-3 hours and carried out at ~$5 \times 10^{-11}$ Torr. We do observe that potassium degrades over time under continuous laser exposure or leaving the sample overnight under vacuum. However, we found that annealing the metasurface to ~150°C for a few minutes, and then depositing another atomic layer of K maintains the superior image quality of the sample, as if K was deposited on freshly prepared metasurface samples.

**Supplementary note: Analysis of PEEM data**

Photoelectron image datasets (*i.e.*, the data cube either as a function of the excitation wavelength or the polarization orientation) were processed to eliminate the impacts of (1) the two-dimensional electron detector's inhomogeneous response, (2) the photoelectron cross section as a function of the excitation wavelength, (3) the image distortion originating from the imperfect alignment of the electron optics, as well as (4) the gradual loss of photoelectron intensity with time that comes from prolonged exposure of the samples to intense visible to near IR pulse laser irradiation. We presume that the heating of the sample due to pulse laser exposure causes the slow K loss, and hence increases the work function of the sample gradually. For (1), we first acquired an image of an area where the sample surface is homogeneous. This reference image was used to correct the inhomogeneous response of the electron detector. For (2), the photoelectron cross section variation due the excitation wavelength was considered by normalizing with the photoelectron intensity variations of the uniform area adjacent to the resonator arrays as a function of the excitation wavelength or the polarization orientation. To eliminate the image distortion (3), we aligned the resonator arrays as if the resonator locations form a perfect orthogonal grid. We found that the distortion within the field of view amounts as much as 20 pixels within a 600 pixel × 600 pixel image (with ~12.5 nm/pixel) depending on the alignment of the electron optics and the position within the detector area. For (4), the gradual loss of the photoelectron intensity was corrected by acquiring two iterations of the wavelength sweep, or of the polarization sweep, and linearly interpolating such that the intensities of the identical features and the baseline in the two consecutive sweeps have the same photoelectron intensity. This process also allows correcting for the sample's spatial drift during the data acquisition.

We note that the PEEM data in this work is presented in the unit of the photoelectron yield. For the visible to near IR excitation data, we applied a square-root to the photoelectron intensity to obtain the photoelectron yield accounting for the two-photon photoemission process.[5] The exposure time of the detector and the laser power are kept constant for all data presented here. The photoelectron yield spectra are averaged over the unit-cell area of the resonator array unless specified otherwise.

We capture multiple photoelectron yield spectra under short exposure and averaged them over the repeating unit-cell area of a single resonator. We note that the resonators near the edges of the arrays are excluded in these averaged spectra owing to the edge effects addressed within the main portion of the manuscript.

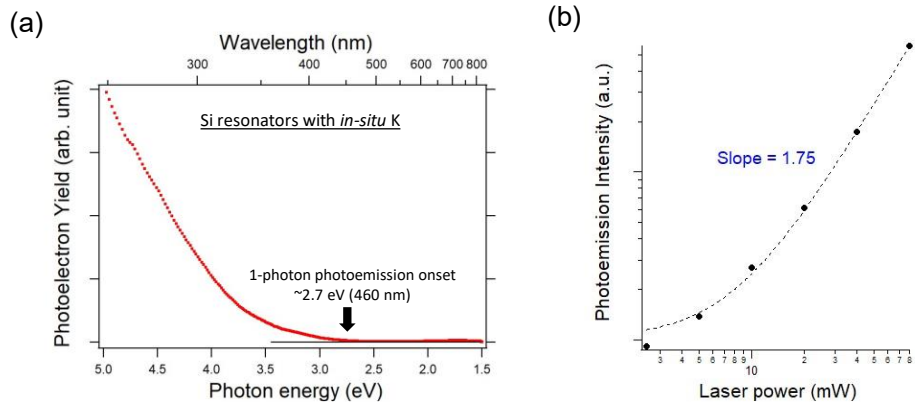

**Supplementary Figure S1**: (a) photoelectron yield spectroscopy of the Si metasurface following *in-situ* K deposition showing the onset of 1-photon photoemission at approximately 2.7 eV. (b) Power dependence of the photoelectron emission at 1.73 eV indicating 2-photon photoemission process.

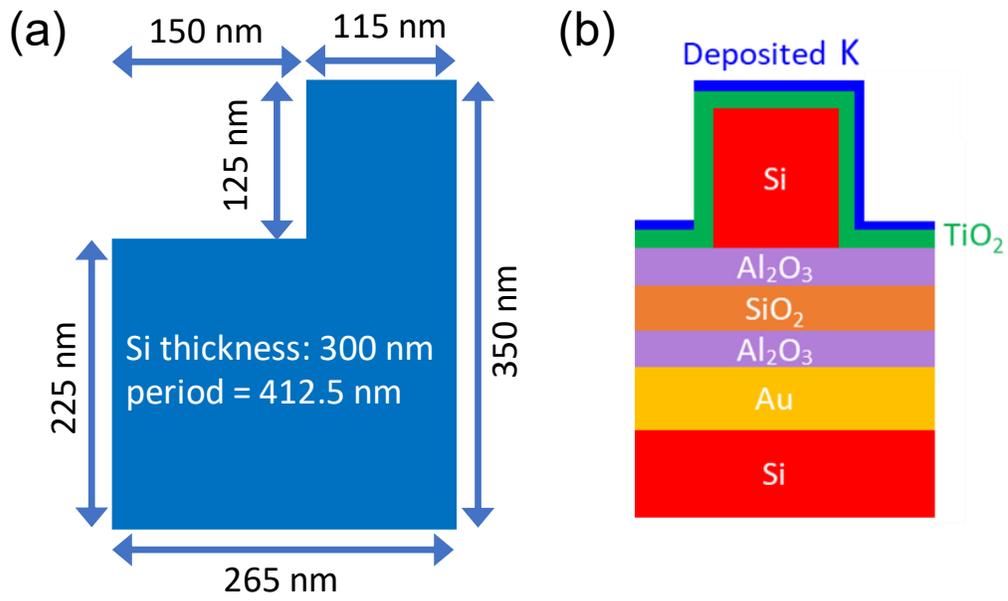

**Supplementary Figure S2**: (a) Detailed geometry of the individual resonator unit that makes-up the Si metasurface (b) 2D cross-sectional schematic of the sample architecture.

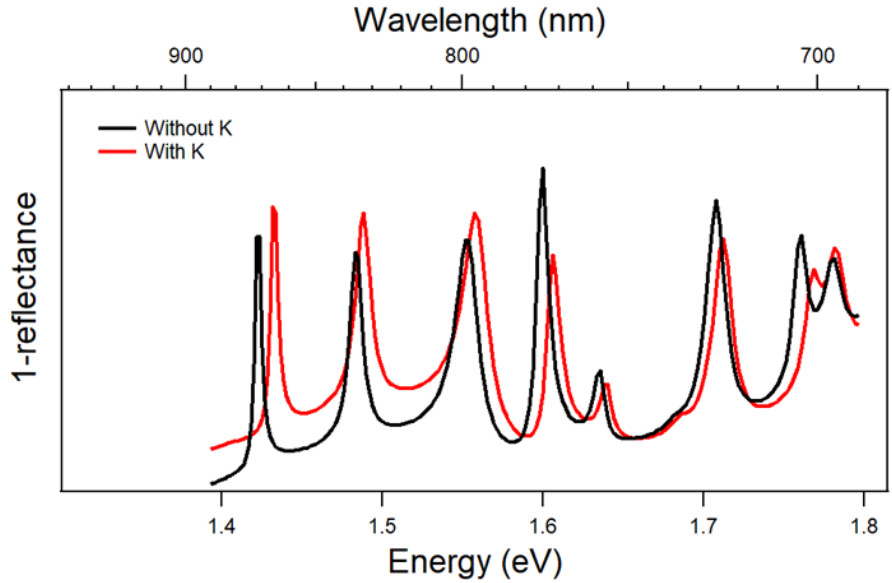

**Supplementary Figure S3**: Effect of K surface layer on simulated spectra. Inverse reflectance spectra (i.e., absorption spectra) of the fabricated Si metasurface calculated using full-wave electromagnetic simulations (COMSOL) with and without a K surface layer for an incident polarization orientation of 90° (see Fig. 2a).

To understand the effect of a K surface layer, simulations were performed using a resonator width of 210 nm, resonator depth of 158 nm, resonator height of 300 nm, notch width of 75 nm, and notch depth of 113 nm. For these simulations, the silicon was treated as non-hydrogenated polycrystalline silicon. The $TiO_2$ layer was 10 nm thick with a conformal K surface layer 2 nm thick. Simulated 1-R spectra are presented in Fig. S3 demonstrating that the K surface layer results in a slight blue shifting of resonances.

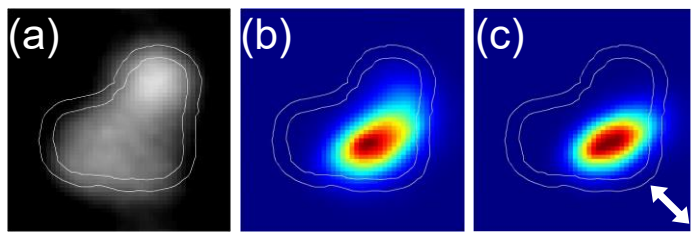

**Supplementary Figure S4**: Two-color illumination imaging [6.44 eV and 1.62 eV (*i.e.*, B-resonance) excitations] showing the resonance location with respect to the resonator geometry. (a) 6.44 eV illumination only. (b) 6.44 eV and 1.62 eV co-illumination. (c) 1.62 eV illumination only. Solid double lines illustrate the approximate locations of the resonator boundaries extracted from (a) and highlight the location of the B-resonance. The polarization for the 1.62 eV illumination is 135° illustrated by the white double end arrow (see Fig. 2 for the definition of the polarization orientation). 6.44 eV illumination (unpolarized) is carried out with an incident angle of 73° off-normal to the sample and using a Xe-lamp with a monochromator.

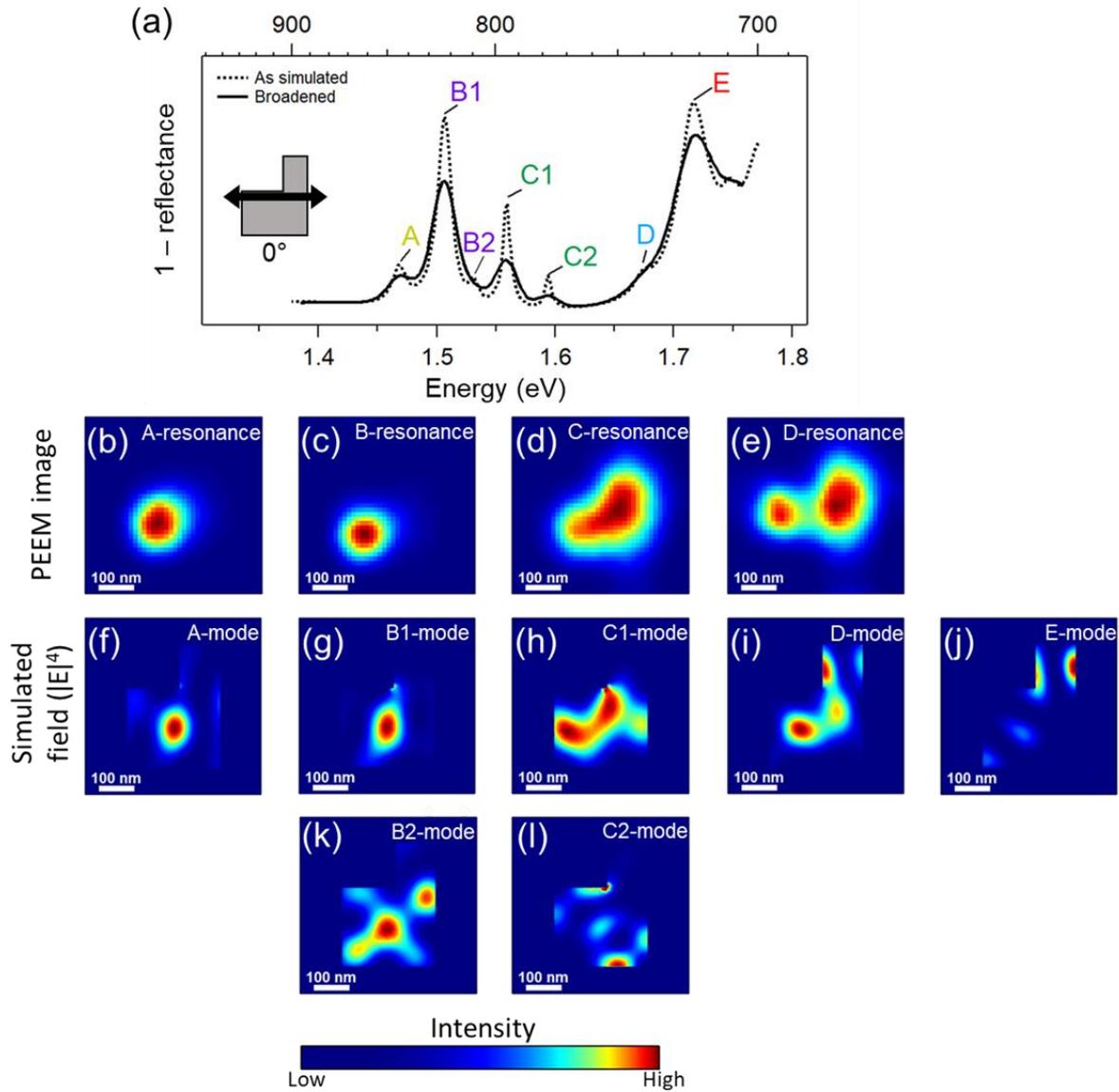

**Supplementary Figure S5**: Excitation energy and polarization dependent nanoscale mode imaging and simulated electromagnetic field profiles of resonators. (a) 1-reflectance spectra of the Si metasurface calculated using full-wave electromagnetic simulations (COMSOL) for an incident polarization orientation of 135°. The dashed line depicts the initial simulated spectra and solid line is the result of broadening with a Gaussian convolution (FWHM = 17.5 meV, ~9 nm). Individual modes in the non-broadened spectra are labeled A to E based on the assignment with experimentally measured resonances as described in the main text. (b), (c), (d), (e) Resonator averaged photoelectron intensity images at resonant excitation energies for a 0° polarization orientation, 1.52 eV (b), 1.62 eV (c), 1.67 eV (d), and 1.73 eV (e). (f), (g), (h), (j), (k), (l) Simulated profiles of the square of the electric field intensity for every mode in (a)

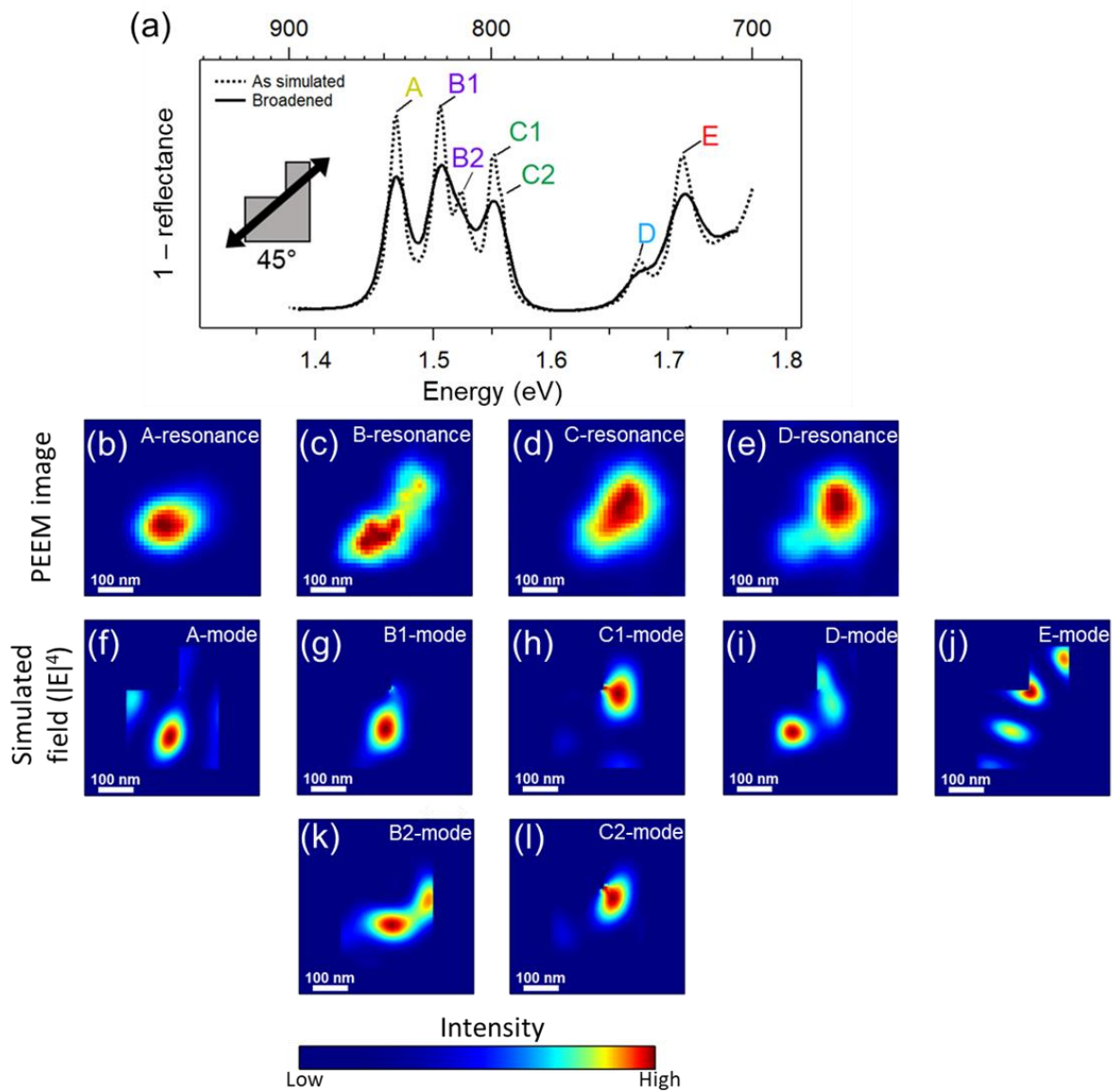

**Supplementary Figure S6:** Excitation energy and polarization dependent nanoscale mode imaging and simulated electromagnetic field profiles of resonators. (a) 1-reflectance spectra of the Si metasurface calculated using full-wave electromagnetic simulations (COMSOL) for an incident polarization orientation of 45°. The dashed line depicts the initial simulated spectra and solid line is the result of broadening with a Gaussian convolution (FWHM = 17.5 meV, ~9 nm). Individual modes in the non-broadened spectra are labeled A to E based on the assignment with experimentally measured resonances as described in the main text. (b), (c), (d), (e) Resonator averaged photoelectron intensity images at resonant excitation energies for a 45° polarization orientation, 1.53 eV (b), 1.61 eV (c), 1.67 eV (d), and 1.75 eV (e). (f), (g), (h), (j), (k), (l) Simulated profiles of the square of the electric field intensity for every mode in (a)

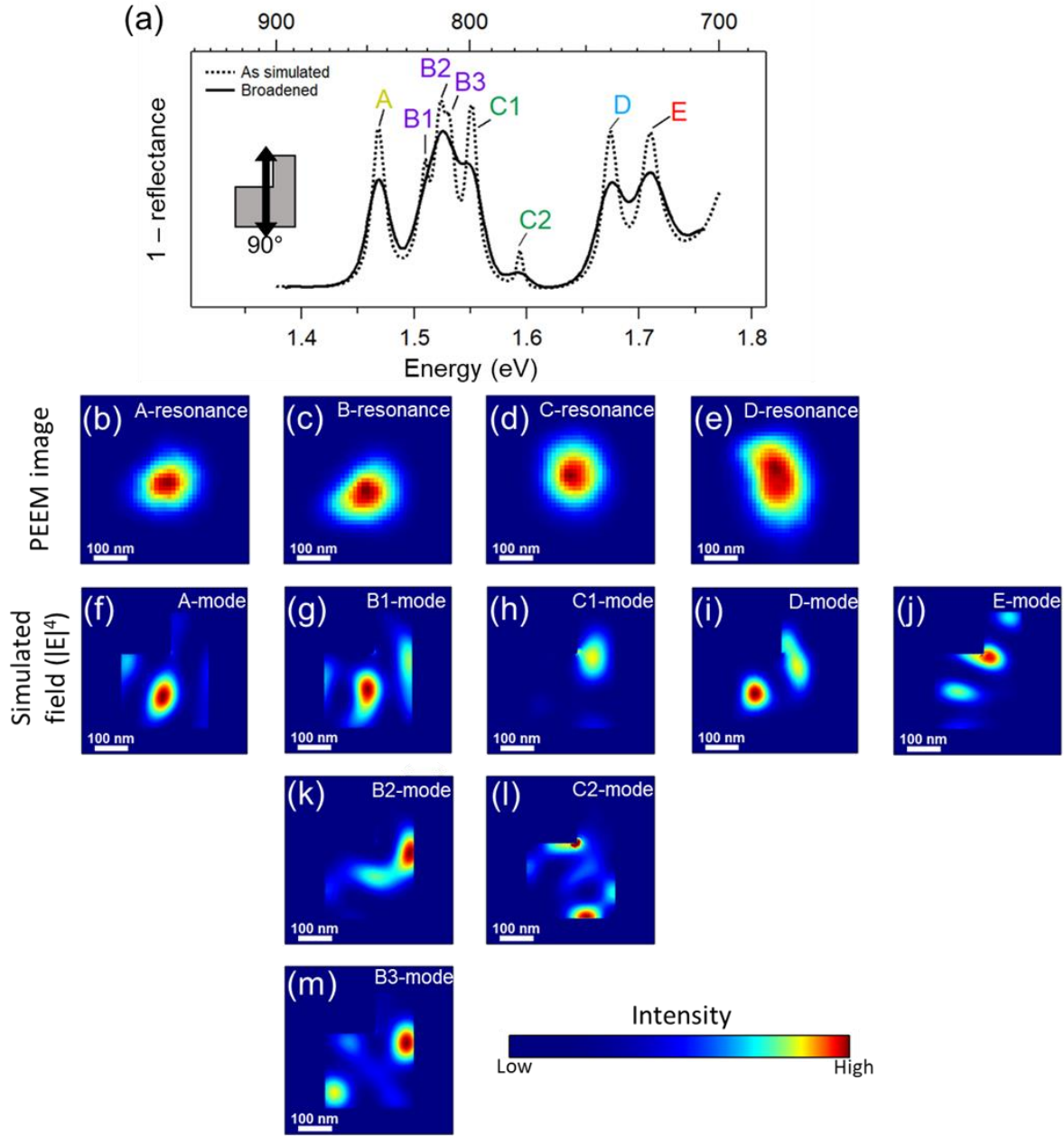

**Supplementary Figure S7:** Excitation energy and polarization dependent nanoscale mode imaging and simulated electromagnetic field profiles of resonators. (a) 1-reflectance spectra of the Si metasurface calculated using full-wave electromagnetic simulations (COMSOL) for an incident polarization orientation of 90°. The dashed line depicts the initial simulated spectra and solid line is the result of broadening with a Gaussian convolution (FWHM = 17.5 meV, ~9 nm). Individual modes in the non-broadened spectra are labeled A to E based on the assignment with experimentally measured resonances as described in the main text. (b), (c), (d), (e) Resonator averaged photoelectron intensity images at resonant excitation energies for a 90° polarization orientation, 1.52 eV (b), 1.62 eV (c), 1.7 eV (d), and 1.73 eV (e). (f), (g), (h), (j), (k), (l), (m) Simulated profiles of the square of the electric field intensity for every mode in (a)

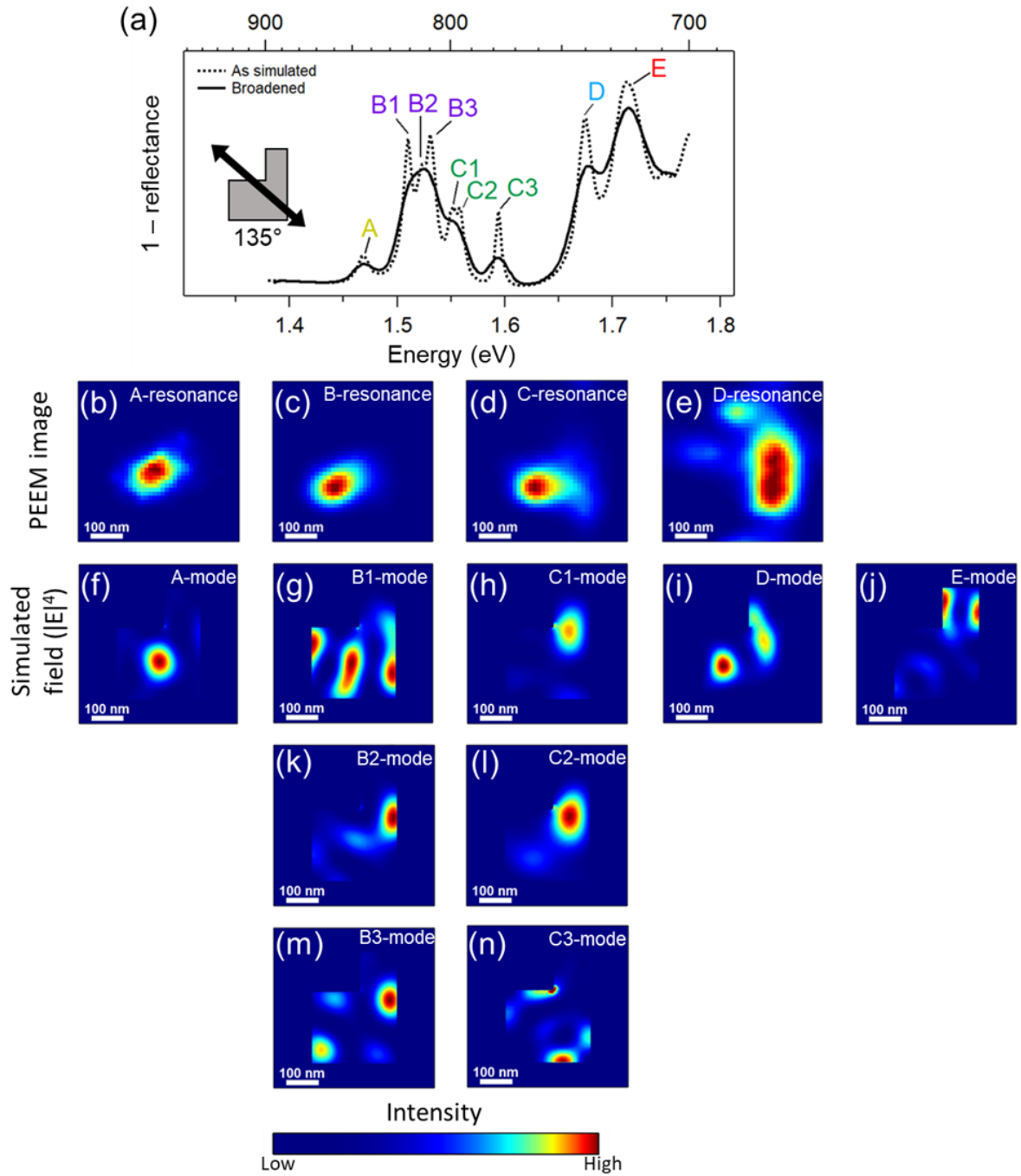

**Supplementary Figure S8:** Excitation energy and polarization dependent nanoscale mode imaging and simulated electromagnetic field profiles of resonators. (a) 1-reflectance spectra of the Si metasurface calculated using full-wave electromagnetic simulations (COMSOL) for an incident polarization orientation of 135°. The dashed line depicts the initial simulated spectra and solid line is the result of broadening with a Gaussian convolution (FWHM = 17.5 meV, ~9 nm). Individual modes in the non-broadened spectra are labeled A to E based on the assignment with experimentally measured resonances as described in the main text. (b), (c), (d), (e) Resonator averaged photoelectron intensity images at resonant excitation energies for a 135° polarization orientation, 1.52 eV (b), 1.62 eV (c), 1.7 eV (d), and 1.73 eV (e). (f), (g), (h), (j), (k), (l), (m), (n) Simulated profiles of the square of the electric field intensity for every mode in (a)

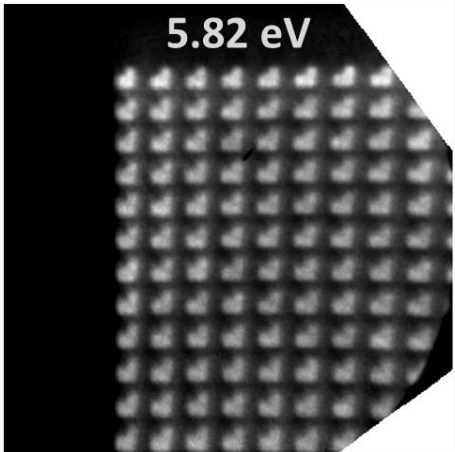
**Supplementary Figure S9:** Photoelectron intensity image of the upper-left corner of a large metasurface array imaged using DUV (5.82 eV) excitation. The region shown is the same as Figure 5 in the main text and Supplementary Figure S10.

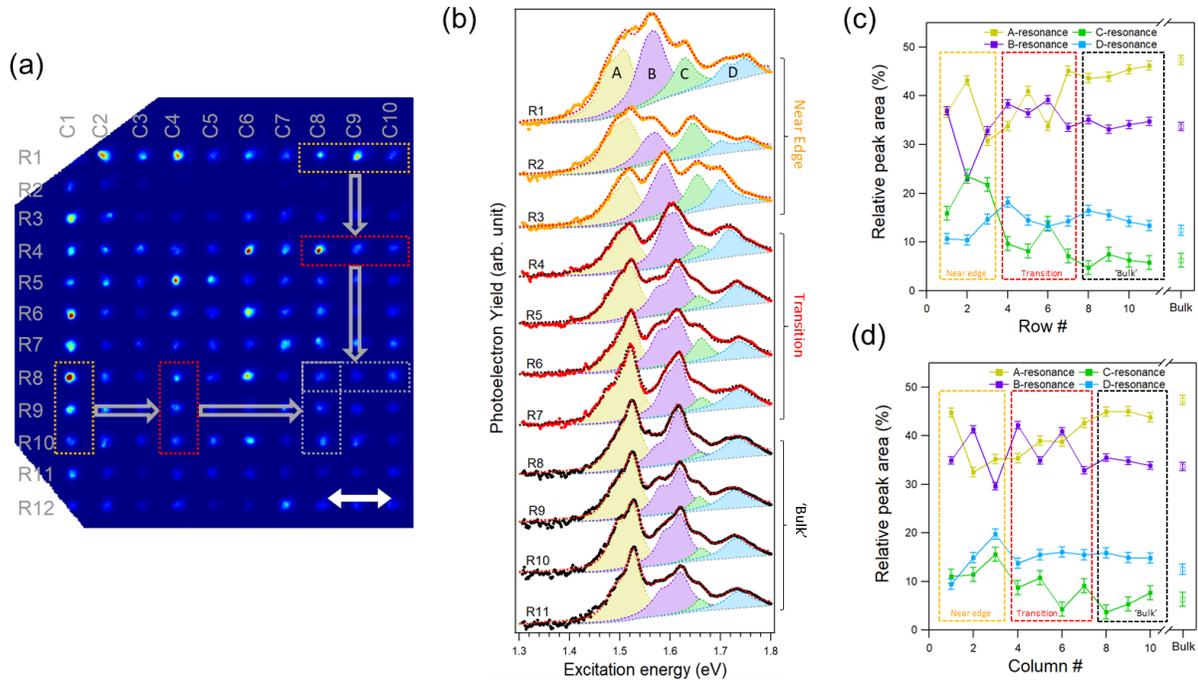

**Supplementary Figure S10**: Evolution of the coherent interactions from the edges to the middle of the resonator array. (a) Photoelectron image of the upper-left corner of a large metasurface array acquired at 1.62 eV excitation (B-resonance) and a 0° polarization orientation (indicated by white double arrow). The rows and columns are designated as R1-R12 and C1-C10 respectively within the image. (b) The area averaged spectra of the resonators as a function of the distance from the left array edge (column number). The spectra are averaged for 3 unit-cells along the vertical direction beginning at R8, as indicated by the yellow, red, and grey doted rectangles. (c, d) The relative peak area for each resonance mode as a function of the column (c) or row (d) number (i.e., distance from left or top array edge respectively). The relative peak areas in (d) come from fits to the spectra averaged over 3 unit-cells along the horizontal direction beginning at C8, as indicated in (a). Relative peak areas from Bulk come from the spectral fits in Figure 2d. Error bars correspond to the standard deviation observed for each mode across the Bulk region averaged for Figure 2d.

[1] Márquez, E.; Blanco, E.; García-Vázquez, C.; Díaz, J. M.; Saugar, E., Spectroscopic ellipsometry study of non-hydrogenated fully amorphous silicon films deposited by room-temperature radio-frequency magnetron sputtering on glass: Influence of the argon pressure. Journal of Non-Crystalline Solids, 2020, 547, 120305.

[2] Berg, M.; Keyshar, K.; Bilgin, I.; Liu, F.; Yamaguchi, H.; Vajtai, R.; Chan, C.; Gupta, G.; Kar, S.; Ajayan, P.; Ohta, T.; Mohite, A. D., Layer dependence of the electronic band alignment of few-layer $MoS_2$ on $SiO_2$ measured using photoemission electron microscopy, Phys. Rev. B 2017, 95, 235406.

[3] Sharma, P. A.; Ohta, T.; Brumbach, M.; Sugar, J. D.; Michael, J. R., Ex Situ Photoelectron Emission Microscopy of Polycrystalline Bismuth and Antimony Telluride Surfaces Exposed to Ambient Oxidation, ACS Applied Materials & Interfaces, 2021, 13, 18218-18226.

[4] Buckanie, N. M.; G̈ohre, J.; Zhou, P.; von der Linde, D.; Horn-von Hoegen, M.; Meyer zu Heringdorf, F.-J., Space charge effects in photoemission electron microscopy using amplified femtosecond laser pulses, J. Phys.: Condens. Matter 2009, 21, 314003.

[5] Stenmark, T.; Könenkamp, R. Photoemission electron microscopy to characterize slow light in a photonic crystal line defect, Phys. Rev B 2019, 99, 205428.